\documentclass[preprint,showpacs,amsmath,amssymb,pra]{revtex4}
\usepackage{dcolumn} 
\usepackage{bm}      
\begin{document}
\title{Remarks on the nature of quantum computation}
\author{Robert Alicki}           \email{fizra@univ.gda.pl}
\affiliation{Institute of Theoretical Physics and Astrophysics,
             University of Gda{\'n}sk, ul. Wita Stwosza 57,
             80-952 Gda{\'n}sk, Poland}
\date{\today}
\begin{abstract}
Two models of computer, a quantum and a classical "chemical machine" designed to compute the relevant part of Shor's factoring algorithm are discussed.
The comparison shows that the basic quantum features believed to be responsible for the exponential speed-up of quantum computations possess their classical counterparts for the hybrid digital-analog computer. It is argued that the measurement errors
which cannot be fully corrected make the computation not efficient for both models.

\end{abstract}
\maketitle

\section{Introduction}
The appearance of Shor's algorithm for factoring integers \cite{shor} beside its potential practical consequences for cryptography
posed a fundamental, even philosophical question: What is the nature of quantum computation process which makes possible efficient
solutions of problems for which the efficient classical algorithms seem not exist?
\par
We remind that the algorithm is efficient if it uses a number of logical steps which is at most polynomial in logarithm of the input size $N$.
\par
From the vast literature on the topic of quantum computations one can extract the following heuristic proposals of the possible answers to
the above question:
\par
A) Quantum superpositions permit quantum computers to perform many computations simultaneously.
\par
B) Entanglement allows to generate and manipulate a physical representation of the correlations between logical entities, without the need to
completely represent the logical entities themselves.
\par
C) Linearity of quantum dynamics makes quantum computation robust against the external perturbations.
\par
D) There exist efficient quantum error correction schemes.
\par
In the course of development of quantum information and computation theory the above statements have been individually challenged by several authors.
For example, in \cite{steane} the picture of "many simultaneous computations" is rebutted, and in \cite{jozsa} the key role of entanglement is questioned.
The parallellism between classical chaos and quantum decoherence, on the essential for quantum computations time scale $\sim\log {\rm dim}{\cal H}$ (${\cal H}$- Hilbert space of the computer), has been
studied in \cite{monte} while the examples of quantum noise which do not satisfy the assumptions of the existing error correction schemes were discussed in
\cite{alhor}.
\par
A quantum computer is a physical system governed by the probabilistic laws of quantum mechanics and  involving both digital entities related to 
discrete spectra of observables and continuous ones associated with arbitrary superpositions and unitary rotations. Therefore a fair comparison demands to consider as
a classical counterpart a probabilistic hybrid digital-analog classical computer. In both cases the analysis of efficiency involves careful estimation of
the used resources (time, space, energy) and the presence of continuous component implies the analysis of errors and their possible corrections.
\par
In the present paper we would like to compare the realizations of the crucial part of the factoring algorithm on a quantum computer with the analogical scheme
performed on a certain hybrid digital-analog "chemical machine". We argue that the typical quantum features A) B) possess their classical macroscopical
counterparts. We discuss also the output measurement errors in both cases, their scaling with the size of the input and the feasibility of corrections.

\section{Factoring algorithm - quantum vs. classical}

The details of the factoring algorithm for a given integer $N$ can be found in \cite{shor,ekert}. Here, we discuss the only essential (in the context of quantum computations) part of this algorithm, namely the computation of the period of the function
\begin{equation}
F_N(a) = y^a {\rm mod}N\ \ ,\ a=0,1,2,...
\label{function} 
\end{equation}
where $1<y<N$ is a randomly choosen fixed integer. From the obvious recurrence formula
\begin{equation}
F_N(a+1) = yF_N(a) {\rm mod}N\ \ ,\ a=0,1,2,...
\label{iteration} 
\end{equation}
it follows that, indeed, $F_N(a)$ is periodic with a period $1<r<N$ and takes $r$ different values. The important fact is that there exists a classical
efficient algorithm for the computation of $F_N(a)$ while an efficient classical algorithm for the computation of $r$ is not known and probably does not exist.
\par
\subsection{Quantum algorithm}
The part of Shor's algorithm used to find the period $r$ consists of four steps performed on a quantum computer consisting of two registers,
the first is a system of $L$ qubits with $q=2^L$ choosen to satisfy $N^2< q < 2N^2$ , the second - $l$ qubits , $N \leq 2^l$ and with an additional computer performing classical efficient "precomputations".
\par
1) Initial state preparation produces the quantum state
\begin{equation}
|\Psi_1> = {\frac{1}{\sqrt{q}}}\sum_{a=0}^{q-1} |a>\otimes |0>
\label{instate} 
\end{equation}
where the state $|a>$ is a product of the qubit states corresponding to the binary representation of $a$.
\par
2)The results of efficient classical precomputation of $F_N(a)$ are encoded into the entangled state
\begin{equation}
|\Psi_2> = {\frac{1}{\sqrt{q}}}\sum_{a=0}^{q-1} |a>\otimes |F_N(a)> \ .
\label{modular} 
\end{equation}
\par
3) Quantum Fourier transform applied to the first register yields
\begin{equation}
|\Psi_3> = {\frac{1}{q}}\sum_{a=0}^{q-1}\sum_{c=0}^{q-1} e^{2\pi iac/q} |c>\otimes |F_N(a)>\ .
\label{fourier} 
\end{equation}
\par
4) Measurements on the first register with respect to the computational basis yield the values of $c$ which allow to find $r$ with sufficiently
high probability by applying auxiliary efficient classical computations. It is important that the number of right outcomes $c$ is equal to $r$ and they
apear with almost equal probabilities.
\par
The main feature of this algorithm is that the quantum operations used in steps 1)-4) can be realized with a sufficient accuracy as compositions of polynomial in $\log N$ number of single-qubit or two-qubit gates (unitary maps). The standard explanation of the fact that this algorithm is exponentially faster than any known classical one refers to the points A) B) in the Introduction. One says, that the superposition in \eqref{instate} allows to compute the function $F_N(a)$ "simultaneously
for $q$ initial values" and the quantum Fourier transform \eqref{fourier}, relying strongly on entaglement, extracts the relevant parameter $r$ without recording the intermediate computation results.

\subsection{Model of chemical computer}

We propose a classical digital-analog machine designed to compute the period $r$ of the function $F_N(a)$. This computation is a kind of Gedankenexperiment which
practical realization may be very difficult but is in a full agreement with the classical laws of thermodynamics. As information carriers we use an ensemble of "polymers" each of them consist of $L$ segments. Any segment can be in two distinguishable states  corresponding to the logical values $0,1$. Therefore the total configuration of a polymer corresponds to a number $a = 0,1,2,...,q-1=2^L-1$ represented by its binary expansion. The thermodynamical
internal energy $E$ is assumed to be independent of the states of segments. The computation of the period $r$ consists of the following steps.
\par
I) Preparation of the initial state which is a thermal equilibrium ensemble of $M$ polymers described by the uniform probability distribution over all configurations $\{a\}$ and the
entropy $S_{in} = k_B ML\log 2$ .
\par
II) Performation of chemical reactions which transform polymer's configuration $\{a\}$ into configuration $\{F_N(a)\}$, $a=0,1,2,...q-1$. It can be done in polynomial in $\log N$
steps involving polynomial in $\log N $ "enzymes" acting locally on polymer segments. The final, nonequilibrium state is a uniform distribution over $r$ configurations with the entropy $S_{out} = k_B M\log r$.
\par
III) Extraction of work during the infinitely slow reversible process of equilibration under isothermal conditions at the temperature $T$. The work is given by a difference
of the free energy $F = E - TS$ and because the internal energy does not depend on the configuration we obtain
\begin{equation}
 W =  T (S_{in} -S_{out}) = k_BT M (L\log 2 - \log r)
\label{work} 
\end{equation}
what allows to find the period $r$ by performing accurate measurements of the work and computing
\begin{equation}
 \log r = L\log 2 - \frac{W}{Mk_BT}\ .
\label{per} 
\end{equation}
The intuitional picture behind this model may be provided by the theory of rubber \cite{vol}.
The elasticity of rubber, consisting of polymers, is entirely due to the entropy, its internal energy is not changed in isothermal processes and its equilibrium state corresponds
to a "random coil". There is also a certain striking similarity to the mechanism of muscle contraction where chemical energy is transformed into mechanical work \cite{bio}. One should also mention in this context the theoretical ideas and experiments on DNA-based computers or more general molecular computations
involving for example protein folding \cite{dna}. In particular, the examples of computations with the macroscopic output measurement are conceptually close to the proposed model.

\subsection{Measurement errors}

Both, presented physical realizations of the computation of the period $r$ are threated by errors of various origin. For the quantum computer irreversible processes of decoherence
and dissipation decrease the probability of obtaining the right result while for the chemical machine due to irreversible processes of the entropy production the equality
\eqref{work} must be replaced by the innequality $W < T(S_{in} - S_{out})$. Ingenious ideas of error correction based on a certain amount of redundancy can improve
the situation. The schemes of quantum error correction should work for certain types of environmental noises\cite{qerror}. For the chemical computer the assumption of infinitely slow process can be relaxed by averaging over an
ensemble of finite-time measurements and using the result of \cite{jar} which in our case gives the relation
\begin{equation}
 \log r = L\log 2 + \log \overline{\exp (-W/Mk_BT)}\ .
\label{jar} 
\end{equation}
Nevertheless, we cannot eliminate all errors. Let us concentrate on the errors in the final measurement of the output. In the quantum case 
we have to perform $L$ independent measurements of the qubit states each of them yields the right result with the probability $2^{-\epsilon}<1$.The sources of errors
in the case of Stern-Gerlach experiment, which is the prototype of an ideal spin measurement, are discussed extensively in \cite{bush}. They are related to the uncertainty of continuous macroscopic parameters like magnetic field direction and position of the spot on the screen and therefore cannot be completely eliminated.
It follows that the probability of correct identification of the output state $|c>$ for the whole register falls down exponentially in $\log N$
\begin{equation}
P(right) = q^{-\epsilon} \sim N^{-2\epsilon}\ .
\label{qer}
\end{equation}
For the classical model we measure the macroscopic work $W$ with an accuracy $\delta W$. Introducing $\gamma = \delta W/Mk_BT$ we obtain the average error
of the period $r$ as $\delta r = \gamma r$. Assuming the Gaussian distribution of errors we can estimate the probability of obtaining the right result as
\begin{equation}
P(right) = \frac{1}{\sqrt{2\pi}\delta r}\sim \frac{1}{\gamma N}
\label{cer}
\end{equation}
which is also exponentially small in $\log N$.
\par
Finaly, we could try to apply error correction schemes to the measurement outcome. Assume for the moment, that in the quantum case for a given input we always obtain a single outcome $c$ in an ideal measurement. Then, by repeating real, unsharp
measurements a sufficient number of times ($\sim \log L$) and using, for instance, majority rule for single-qubit outcomes we could effectively recover the right state. However,
as mentioned in Section IIA the number of right outcomes $c$ is equal to $r$ and therefore exponentially large what makes this method not applicable.
For the classical model repeating measurement $K$ times can reduce an error by a factor $1/{\sqrt K}$ what is also an inefficient procedure.

\section{Conclusions}

We present, on the level of Gedankenexperiment, the comparison between the quantum computer and the chemical one, both designed to execute the crucial part of the factoring algorithm. The analysis shows that the typical quantum effects, which are believed to allow computations beyond the scope of any classical computer, possess their classical counterparts. The initial uniform superposition of quantum inputs corresponds to a random ensemble of classical ones. The possibility of extracting
the final result of computation without recording the intermediate steps seems to be the feature of analog computations involving continuous variables and present in both examples. This advantage of analog computers is overshadowed by their vulnerability to errors. We show on the example of the final measurement that the assymptotic
behavior of errors makes the computation inefficient for both models. The error correction of the output measurement is also not efficient for quantum and classical case as well.
\par
It seems that a quantum computer can be treated as a kind of hybrid digital-analog computer with all its advantages and drawbacks. Digital-analog machines were quite successful
in the past and probably dominate information processing in living organisms despite their bad scaling of errors. Therefore, one cannot exclude that the future
technology will allow quantum computers to be useful for certain specific tasks.
\begin{acknowledgments}
Discussions with Micha\l\ , Pawe\l\ and Ryszard Horodecki's are gratefully acknowledged.
\end{acknowledgments}

\end{document}